\documentclass[prl,twocolumn,showpacs,superscriptaddress]{revtex4}
\usepackage{graphicx,amssymb,amsmath,psfrag,color}
\begin{document}
\definecolor{darkgreen}{rgb}{0,0.5,0}

\title{Crossover from adiabatic to sudden interaction quench in a Luttinger liquid}

\author{Bal\'azs D\'ora}
\email{dora@kapica.phy.bme.hu}
\affiliation{Department of Physics, Budapest University of Technology and
  Economics, Budafoki \'ut 8, 1111 Budapest, Hungary}
\author{Masudul Haque}
\affiliation{Max-Planck-Institut f\"ur Physik komplexer Systeme, N\"othnitzer Str. 38, 01187 Dresden, Germany}
\author{Gergely Zar\'and}
\affiliation{Department of Theoretical Physics, Budapest University of Technology and Economics, H-1521 Budapest, Hungary}

\date{\today}

\begin{abstract}
Motivated by recent experiments on interacting cold atoms, 
we analyze interaction quenches in Luttinger liquids (LL), where the
interaction is ramped from zero to a finite value within a finite time. 
The fermionic
single particle density matrix reveals several regions of spatial and temporal
coordinates relative to the quench time, termed as Fermi liquid,
sudden quench LL, adiabatic LL regimes, and a LL regime with
time dependent exponent. The various regimes can also be observed
in the momentum distribution of the fermions, directly accessible 
through time of flight experiments. Most of our results apply to arbitrary
quench protocols.  
\end{abstract}


\maketitle

Non-equilibrium dynamics and strong-correlation phenomena in quantum many body
systems are topics at the forefront of contemporary physics.
%
%
When these two fields are combined, namely when strongly correlated systems
are driven out-of-equilibrium, we face a real challenge.  Experimental
advances on ultracold atoms \cite{BlochDalibardZwerger_RMP08} have made the
time dependent evolution and detection of quantum many-body systems possible,
and in particular, quantum quenching the interactions by means of a
 Feshbach resonance or time dependent lattice parameters has triggered enormous
theoretical \cite{calabrese,rigol,dziarmagareview,polkovnikovrmp} and
experimental \cite{hofferberth,haller,kinoshita,paredes} activity.

Luttinger liquids (LL) are
ubiquitous as effective low-energy descriptions of gapless
phases in various one-dimensional (1D) interacting
systems~\cite{giamarchi,nersesyan}.  In 1D fermions, e.g., 
Landau's Fermi liquid description breaks down for any finite
interaction, and the low-energy physics is described by bosonic
collective modes with linear dispersion, and is characterized by anomalous
non-integer power-law dependences of correlation functions.  
The LL similarly arises as the low-energy description of interacting 
1D bosons or that of spin chains~\cite{giamarchi}.  

A LL could also be driven out of
equilibrium through transport \cite{gutman, perfetto}, e.g.,
but here we shall concentrate on time dependent 
changes of the interaction parameter, which is of particular relevance
for cold atomic systems.
Sudden quenches (SQ) of the interaction in LLs have been considered recently
by several authors \cite{cazalillaprl,uhrig,iucci},  
and the idea of the other extreme limit of
adiabatic parameter ramps is often invoked, too.  
However,  experimental ramps cannot take infinite time, and are not 
instantaneous, either. Here we study, how a nonzero
quench time, $\tau\neq0$, influences the final state of the system
after a quantum quench. As we show, a finite $\tau$ 
leads to `heating' effects, and generates excitations in the final state.   
Moreover, it  amounts in the appearance of additional
energy ($\sim 1/\tau$) and corresponding length scales: 
while in certain space-time regions the system 
reveals  universal near-equilibrium (adiabatic) correlations
\cite{polkovnikovrmp}, in other regimes  
renormalized Fermi liquid (FL) or sudden quench (SQ) 
behavior is found.

Thus motivated, let us study the LL Hamiltonian\cite{giamarchi}
\begin{equation}
H=\sum_{q\neq 0} \omega(q)b_q^\dagger b_q
+\frac{g(q,t)}{2}[b_qb_{-q}+b_q^+b_{-q}^+],
\label{hamilton}
\end{equation}
with $\omega(q)=v|q|$ ($v$ being the bare "sound velocity"), 
and $b_q^\dagger$ the creation operator of a bosonic density wave. 
The interaction $g$ is changed 
 from zero to a nonzero value within a quench time $\tau$, 
 $g(q,t)=g_2(q)|q|Q(t)$, with 
$Q(t)$ encoding the explicit quench protocol, and satisfying
$Q(t> \tau)= 1$ and $Q(t<0)=0$ ~\footnote{To avoid instabilities, $v>|g_2(q)|$ is assumed.}.
In particular, for a linear quench, 
$Q(t)=t\Theta(t(\tau-t))/\tau+\Theta(t-\tau)$ with $\Theta(t)$ the Heaviside functions.
This setting is very general, and equally applies to switching on  the
interaction in a  spinless fermion system,
 quenching away from the
Tonks-Girardeau limit of a 1D Bose gas~\cite{paredes}, or turning 
on the $Z$-component of the interaction in an
$XXZ$ spin chain\cite{giamarchi}. To make contact with 
previous work \cite{cazalillaprl,uhrig,iucci},
here we focus on 
fermionic correlators in detail, but most of our results 
can be trivially generalized to interacting bosonic systems\cite{dora}.

We describe  time-evolution  using the Heisenberg
equation of motion, leading to
\begin{gather}
i\partial_t b_q= [b_q,H]= \omega(q)\;b_q+g(q,t)\;b^+_{-q}, 
\label{heisenberg}
\end{gather}
and similarly, $i\partial_t b^+_{-q}=-\omega(q)b^+_{-q}-g(q,t)b_{q}$.
Solutions of these are of the form
\begin{gather}
b_q(t)= u(q,t)\;b_q(0) +v^*(q,t)\;b^+_{-q}(0)\;,
\label{timedepop}
\end{gather}
where all the time dependence is carried by the prefactors,
 $u(q,t)$ and $v(q,t)$, and the operators 
on the r.h.s. refer to non-interacting bosons 
before the quench. All expectation values are thus taken in terms of
the initial density matrix of the latter (or vacuum at $T=0$). 
The bosonic nature of the quasiparticles requires $|u(q,t)|^2-|v(q,t)|^2=1$. 
From Eqs. \eqref{heisenberg}-\eqref{timedepop}, we obtain
\begin{gather}
i\partial_t\left[\begin{array}{c}
u(q,t)\\
v(q,t)\end{array}\right]=\left[\begin{array}{cc}
\omega(q) & g(q,t)\\
-g(q,t) & -\omega(q)
\end{array}\right]
\left[\begin{array}{c}
u(q,t)\\
v(q,t)\end{array}\right],
\label{beq}
\end{gather}
with the initial condition $u(q,0)=1$, $v(q,0)=0$.
Since both $\omega(q)$ and $g(q,t)$ are even functions of $q$,
$u(q,t)$ and $v(q,t)$ must be so, too. 
By Eq.~\eqref{beq}, all time dependence has been transferred 
to the Bogoliubov coefficients, and therefore expectation values of the
time dependent bosonic modes and non-equilibrium dynamics 
are calculable using standard
techniques developed for equilibrium \cite{giamarchi},
once the solutions of Eq.~\eqref{beq} are known.

Before discussing a continuous quench, let's see how limiting cases
are recovered from Eq. \eqref{beq}. 
The adiabatic limit follows from replacing $g(q,t)$ with its time
independent final value, and looking  
for the stationary solutions of Eq. \eqref{beq} at a given energy
while ignoring the initial conditions. 
The SQ limit requires only the replacement of   $g(q,t)$ by its final,
time independent value, and solving the resulting linear 
differential equation with the initial conditions satisfied.

With a linear quench, Eq. \eqref{beq} realizes the
non-hermitian Landau-Zener model \cite{nonhermitianlz}, which can be
solved exactly in terms of the parabolic cylinder function. 
However, the exact solution does not yield an immediate and transparent
physical picture.  Therefore, to obtain more insight, 
we assume that $g(q,t)\ll \omega(q)$ (i.e., $g_2(q)\ll v$) 
for all $t$ and $q$, and solve Eq.~\eqref{beq} perturbatively in the
interaction. To lowest order in $g_2(q)$, we obtain
$u(q,t)\approx\exp(-i\omega(q)t)$ and
\begin{gather}
v(q,t>0)\approx i\int_0^tdt'g(q,t')\exp(i\omega(q)(t-2t'))\;.
\label{eq:v_pert}
\end{gather}
Higher order corrections to 
$u(q,t)$ and $v(q,t)$ are of order $g_2^2(q)$ and $g_2^3(q)$,
respectively.
 We have also checked numerically that  Eq.~\eqref{eq:v_pert} is 
indeed applicable for any $t$ and $\tau$, as long as
$g_2(q)\ll v$ ~\footnote{The solution can easily be extended to include higher powers of
$g_2(q)$ for a linear quench.}.
In the SQ ($\tau\rightarrow 0$) and adiabatic ($\tau \to \infty$)
limits we obtain 
\begin{gather}
v(q,t>\tau)\approx\frac{g_2(q)}{2v}\times \left\{\begin{array}{cc}
2i\sin(\omega(q)t) & \textmd{for } \tau\rightarrow 0,\\
-\exp(-i\omega(q)t) & \textmd{for } \tau\rightarrow \infty,
\end{array}\right.
\end{gather}
 reproducing to lowest order in $g_2(q)$ the SQ results \cite{cazalillaprl,iucci} and the equilibrium
Bogoliubov transformation \cite{solyom,giamarchi}, 
respectively.


We are now in position  to obtain information about
physical observables.  We start with the evolution of the total energy of the
system.
We take the energy of our initial vacuum  state to be 
zero.  In the fermionic setting,  this
corresponds to measuring the energy with respect to the 
energy of the non-interacting  Fermi sea. 
The expectation value of Eq. \eqref{hamilton} is then evaluated in the
Heisenberg picture, where the expectation value is taken with
the non-interacting ground state, and $v(q,t)$ and $u(q,t)$ 
as obtained from Eq. \eqref{timedepop}
keeping track of the time evolution of the system.  We 
thus obtain for $ \langle H(t)\rangle$ 
after the quench ($t>0$)
\begin{gather}
\langle H\rangle=\sum_{q\neq0}\omega(q)n_B(q)+(2n_B(q)+1)\textmd{Im}[v^*(q,t)\partial_tv(q,t)]\;,
\nonumber
\end{gather}
with $n_B(q)=1/(\exp(\omega(q)/T)-1)$ the Bose function.
The expression above is time independent for $t>\tau$, as expected. 
At $T=0$ and $t>\tau$, and an interaction of finite range, 
 $g_2(q)=g_2 \exp(-R_0 |q|/2)$, we obtain 
\begin{gather}
\langle H \rangle=E_{gs}\left[1-
\left(\frac{\tau_0}{\tau}\right)^2
\ln\left(1+\left(\frac{\tau}{\tau_0}\right)^2\right)
\right]\;
\end{gather}
for a linear quench.
Here we introduced the
microscopic time scale, $\tau_0 \equiv R_0/2 v$, and 
$E_{gs}=-Lg_2^2/4\pi vR_0^2$ is the adiabatic  ground
state energy shift to lowest order in $g_2$, with $L$ the system
size.  The second term corresponds to quasiparticle excitations 
resulting from the finite quench speed. 
In the SQ limit, $\tau\ll \tau_0$,
 the energy of the system is only slightly shifted\cite{bernier}, $\langle H\rangle=E_{gs}(\tau/\tau_0)^2/2$.  
This holds true for a general quench, i.e. $\langle H\rangle\sim (\tau/\tau_0)^2$ when
$\tau\rightarrow 0$ with a quench dependent coefficient.
In the adiabatic
limit, $\tau\gg\tau_0$, on the other hand, 
the excess energy (or ``heating'') vanishes as
$-2E_{gs}\ln({\tau}/{\tau_0})\;{\tau_0^2}/{\tau^2}$ in accord with the so-called analytic response of Ref. \cite{polkovnikovnatphys}.
This remains valid for general smooth quenches displaying kink(s) (discontinuity in the derivative) and bounded $\partial_tQ(t)$.
Smooth quenches without kinks but with bounded $\partial_tQ(t)$ produce also a universal decay as $\sim 1/\tau^2$, while
the $\tau$ dependence of the heating becomes non-universal for protocols with a diverging $\partial_tQ(t)$\cite{degrandi}.
The
crossover between the SQ and adiabatic limits occurs when
$\tau\sim \tau_0$, which typically 
translates to $\tau\sim 1/J$ in an optical lattice, 
with  $J$  the hopping
integral in the underlying microscopic lattice Hamiltonian.

In the fermionic context, the structure of the non-equilibrium 
dynamics can be well demonstrated by means of 
the fermionic one-particle density matrix. Since the fermion field 
decomposes to right-going and a left-going parts, $\Psi(x)= 
e^{ik_F x}\Psi_r(x) +e^{-ik_F x}\Psi_l(x)$, it is enough to concentrate 
on the right-going part of the density matrix, 
$$
G_r(x,t)\equiv \langle \Psi_r^+(x,t)\Psi_r(0,t)\rangle\;,
$$ 
describing excitations  around the right Fermi momentum, $k\approx k_F$. 
The right-going field,  $\Psi_r(x)$, can be expressed in terms of the
LL bosons as \cite{giamarchi}
\begin{gather}
\Psi_r(x)=\frac{\eta_r}{\sqrt{2\pi\alpha}}\exp\left(i\phi_r(x)\right) \;,
\end{gather}
where $\eta_r$ denotes the Klein factor, 
and $\phi_r(x)=\sum_{q>0}\sqrt{2\pi/
  |q|L}e^{iqx-\alpha |q|/2}b_q+h.c.$.
Following standard steps~\cite{giamarchi,nersesyan}, we obtain then
\begin{gather}
G_r(x,t)=G^{0}_r(x)\exp\left(-\sum_{q>0}\left(\frac{2\pi}{qL}\right)4\sin^2\left(\frac{qx}{2}\right)\times\right.\nonumber\\
\left.\times
\left[n_B(q)+|v(q,t)^2|(2n_B(q)+1)\right]\right),
\label{greenboson}
\end{gather}
where $G^{0}_r(x)={i}/({2\pi(x+i\alpha)})$ denotes the free fermion propagator, 
with $\alpha$ an ultraviolet regulator.
At $T=0$, the Bose functions vanish, and only $v(q,t)$, i.e. the mixing between $b_q$ and $b_{-q}^+$ determines the dynamics.
%
%

Let us first discuss the properties of $G_r(x,t)$
long after the quench, $t\gg \tau$. 
In this limit, we can show \cite{EPAPS} that, 
independently of the quench protocol, $Q(t)$, 
the one-particle density matrix exhibits universal properties, 
\begin{gather}
\frac{G_{r}(x,t)}{G^{0}_r(x)}\sim 
\left\{
\begin{array}{ll}
A\left(\tau /{\tau_0}\right)
\left(\dfrac{R_0}{\min \{|x|,2vt\}}\right)^{\gamma_{\rm SQ}} & \hspace*{-4mm}\textmd{for }
|x|\gg 2v\tau,
\\
\left(\dfrac{R_0}{|x|}\right)^{\gamma_{\rm ad}}
 & \hspace*{-4mm}\textmd{for } |x|\ll 2v\tau,
\end{array}
\right.
\label{largex}
\end{gather}
where $\gamma_{\rm SQ}= {g_2^2}/{v^2}+\dots$ and  
$\gamma_{\rm ad}= {g_2^2}/{2 v^2} + \dots$
denote the perturbative sudden quench and adiabatic exponents, 
respectively.  The prefactor $A\left(\tau /{\tau_0}\right)$ depends on 
the speed of the quench: For a sudden quench
it is $A(\tau\ll\tau_0)\sim 1$, while for slower quenches $A(\tau>\tau_0)\sim
(\tau/\tau_0)^{\gamma_{ad}}$. 

\begin{figure}[t]
\centering
{\includegraphics[width=8.3cm,height=4.1cm]{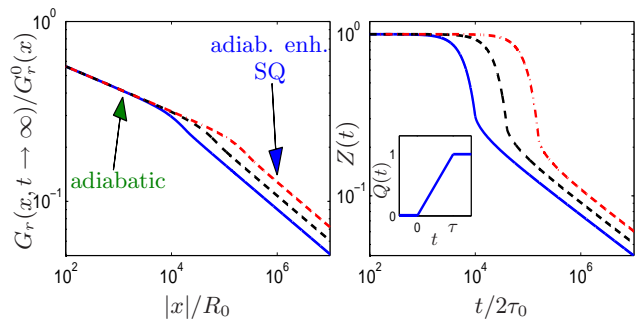}}
\caption{(Color online) Left: the long time ($\tau\ll t\rightarrow\infty$),
steady state limit of the one-particle density matrix is plotted on loglog
scale for a linear quench and $2g_2=v$ as a function of $|x|$, exhibiting the crossover from adiabatic behavior
(lower line in Eq. \eqref{largex}) at small $|x|$ to SQ behavior with
adiabatic enhancement (upper line in Eq. \eqref{largex}) at large $|x|$.  The curves are
plotted for $\tau/2\tau_0=10^4$, $4\times 10^4$ and $16\times 10^4$ from bottom
to top in both panels.  Right: Landau's quasiparticle weight, $Z(t)$ is
plotted on a loglog scale as a function of $t$, bridging between the weakly
interacting Fermi liquid to strongly suppressed $Z\ll 1$ with adiabatic
enhancement (Eq. \eqref{eqtilde1}).  Inset: the linear quench protocol is shown.
\label{figdensitymatrix}}
\end{figure}

Thus even for $t\to \infty$, 
instead of one single power-law, $G_r$
interpolates between the SQ and adiabatic limits. 
This is shown in Fig.~\ref{figdensitymatrix} for a linear quench\cite{EPAPS}.
Physically, it is easy to understand the cross-over behavior observed in 
$G_r$: 
a finite-time  quench is experienced by slow excitations of energy  
$\omega(q)<1/\tau$  as a sudden change, while fast excitations with 
 $\omega(q)>1/\tau$ can adjust to the change in the interaction
strength adiabatically. Since high (small) energy excitations
determine the short (long) distance correlations, the tail 
of $G_r$ is governed by the SQ exponent~\cite{cazalillaprl}, 
while the short distance behavior is described by the adiabatic 
exponent. It is remarkable that  
for slow enough quenches, $\tau\gg \tau_0$, 
the quench time manifests itself explicitly through  
an adiabatically \textit{enhanced} prefactor 
 $A\sim (\tau/\tau_0)^{\gamma_{\rm ad}}$ of the asymptotic tail 
 as also shown in Fig.~\ref{figdensitymatrix}. 
Thus while the spatial decay of Eq. \eqref{largex} contains the
SQ exponent, its $\tau$ dependence reveals the adiabatic LL exponent. 
 For a finite $t\gg\tau$ but $2vt\ll |x|$, $G_r(x,t)$ decays 
asymptotically 
as $i Z(t)/2\pi x$, with a finite quasi-particle weight, 
\begin{gather}
Z(t\gg\tau,\tau_0)\sim A(\tau/\tau_0)\; \left(\frac{\tau_0} t\right)^{\gamma_{\rm SQ}}.
\label{eqtilde1}
\end{gather}
Thus the exponent observed in $Z(t)$
is identically $\gamma_{\rm SQ}$ for $t\gg \tau$, but the 
finite quench time amounts in a 
 quasiparticle weight increased by a factor,
  $A\sim (\tau/\tau_0)^{\gamma_{\rm ad}}$ for $\tau\gg \tau_0$.
Although these results were obtained perturbatively, they carry over
to the non-perturbative limit, too, with the only exception that 
the exponents $\gamma_{\rm SQ}$ and $\gamma_{\rm ad}$ must be replaced
by their exact value.

All these spatial features appear also in the time-dependent momentum
distribution of the fermions, $n(k,t)$, directly measurable through 
time of flight experiments. In particular, at $T=0$ and finite $t\gg \tau$, 
$n(k,t)$ exhibits a jump of size $\sim Z(t)$ at $k=k_F$, while it
approximately scales for $|\tilde k| \gg 1/2vt$ as 
\begin{gather}
n(k)-\frac 12 \sim -\textmd{sign}(\tilde k)\times\left\{\begin{array}{ll}
A(\tau/\tau_0)\; |\tilde k R_0|^{\gamma_{\rm SQ}}, & |\tilde k|\ll
\frac1 {2v\tau}\;,
\\
|\tilde k R_0|^{\gamma_{\rm ad}},  & |\tilde k|\gg
\frac 1 {2v\tau},
\end{array}\right.
\label{mdfermion}
\end{gather}
for $\tilde k \equiv k-k_F$, $|\tilde k|\ll k_F$, and $t\gg\tau$. 
Thus the time scale of the quench is also imprinted in the momentum
distribution, which also shows a cross-over behavior
between the SQ and the adiabatic limits.  
For adiabatic quenches, $\tau\rightarrow\infty$, 
we recover the equilibrium LL exponent, while close to $k_F$, the
momentum distribution  is
enhanced by a  factor $A(\tau/\tau_0)$ compared to 
the SQ behavior\cite{cazalillaprl,iucci}.

The above analysis can be extended to the short time region,
$t\ll \tau$, where the behavior found depends explicitly on the 
quench protocol\cite{EPAPS} as
\begin{equation}
{G_{r}(x,t)}\sim {G^{0}_r(x)}\left(\frac{R_0}{\min\{|x|,2vt\}}\right)^{\gamma(t)},
\label{fg2}
\end{equation}
where $\gamma(t) ={g_2^2Q^2(t)}/{2v^2} + \dots$.
For short distances, $|x|\ll 2v t$, the spatial correlations decay with a time-dependent exponent, 
and this  region  can thus be  characterized as
 a weakly interacting LL (t-LL). For $|x|\gg 2v t$, on the other hand, 
similar to $t\gg \tau$, correlations remain almost unaffected  by
interaction, and a Fermi liquid regime is found. 
 For  $t\ll \tau_0$, $Z(t)\simeq 1$ 
as in the initial Fermi gas, but for $t\gg \tau_0$ we 
 recover a Fermi liquid behavior with a 
reduced quasiparticle weight as
\begin{gather}
Z(\tau_0\ll t\ll \tau )\sim 
\left(\dfrac{\tau_0}{ t}\right)^{\gamma(t)}
\label{fg1}
\end{gather}

\begin{figure}[h!]
{\includegraphics[width=6cm]{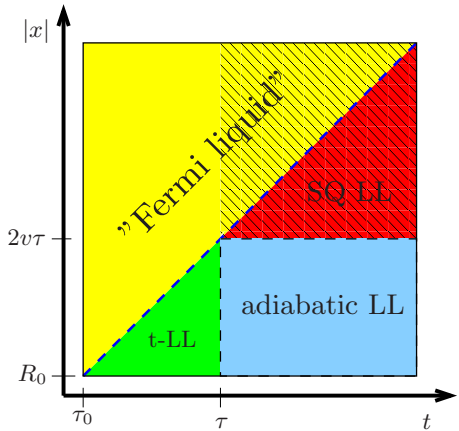}}
\caption{(Color online) 
The schematic universal spatial-temporal characteristics of a quenched LL,
with the boundaries denoting \textit{crossovers}. 
In the adiabatic LL regime, the LL exponent of the final state, 
$\gamma_{\rm ad}$ governs spatial correlations, while in the 
SQ LL region, correlations decay with the SQ exponent, 
$\gamma_{\rm SQ}>\gamma_{\rm ad}$. Correlations are adiabatically
increased
by an amplitude, 
$A\sim (\tau/\tau_0)^{\gamma_{\rm ad}}$ in the shaded region. 
 In the  Fermi liquid region 
a time dependent quasiparticle residue is found (see Eq.~\eqref{fg1}),
 while in the time-dependent LL (t-LL) region a quench protocol-dependent
weakly interacting LL is found with a time dependent exponent, 
Eq. \eqref{fg2}. 
The dashed line denotes $|x|=2vt$, i.e. the light-cone\cite{calabrese}.
For $\tau\ll \tau_0$, the SQ physics of Ref.~\cite{cazalillaprl,iucci}  
dominates everywhere.
\label{figphase}}
\end{figure}

The quasiparticle weight thus slowly decreases during the quench, 
and excitations 
remain similar to those in the initial Fermi gas with a reduced weight
($Z<1$) for $t<\tau$. After the quench, $t>\tau$, 
the quasiparticle weight continues
to  decrease as  a power-law, and it resembles to an interacting
(heavy) Fermi  liquid with $1/x$ spatial decay and $Z\ll 1$ 
quasiparticle residue, Eq. \eqref{eqtilde1},  as was also found for sudden 
quenches~\cite{uhrig,iucci}. This situation is shown in detail 
in Fig.~\ref{figdensitymatrix}, where $Z(t)$ is plotted for the
special case of a  linear quench\cite{EPAPS}. 
Our results are summarized in  Fig. \ref{figphase}.

The main effect of finite temperatures ($T>0$) in Eq. \eqref{greenboson} is to in\-tro\-du\-ce
 a new 
time or length scale, $1/T$ or $v/T$, respectively, above which
the power law behaviour of LL disappears, and gives way to exponentially 
suppressed behaviour as $\exp(-cT\textmd{max}[|x|/v,t])$ with $c$ some
constant.  
Our findings in Fig. \ref{figphase} survive 
in the region $(t,\tau,|x|/v)<1/T$.

Ultracold fermionic gases have been realized using several atoms such as
$^{40}$K\cite{moritz,gunter,kohl}, $^6$Li\cite{liao},
$^{171}$Yb-$^{173}$Yb\cite{fukuhara}, $^{163}$Dy\cite{lu}  
and $^{87}$Sr\cite{desalvo}, and
temperatures well in the quantum degeneracy regime were 
reached ($T<0.1 \;E_F$, with $E_F$ the Fermi energy). 
 All these atomic systems feature tunable interactions, 
essential to address quench dynamics. 
Among these, 1D configurations have been realized 
using $^{40}$K\cite{moritz,gunter}, $^6$Li\cite{liao}, and 
the momentum distribution has been measured in time of flight (ToF)
experiments in 2D\cite{desalvo} and 3D\cite{kohl,fukuhara} Fermi
gases. Therefore, by applying ToF imaging or momentum resolved rf
spectroscopy\cite{moritz}, 
the observation of the momentum distribution 
of Eq.~\eqref{mdfermion}  is  within reach for 1D fermions~\cite{moritz}. 
Furthermore, the specific momentum distribution of a 
LL has already been observed in the Tonks-Girardeau limit of 1D Bose 
systems\cite{paredes}, which exhibit
fermionic properties in this strongly interacting regime, 
and reveal after an interaction quench
features similar to the ones found for 
fermions~\cite{dora}.

In summary, we have studied continuous interaction quenches in LL, bridging smoothly between the SQ and adiabatic limits.
The resulting dynamics is largely influenced by the finite quench time for fermions, and in particular, the momentum 
distribution exhibits a crossover from the adiabatic LL to that of the SQ with the extra adiabatic enhancement factor $(\tau/\tau_0)^{\gamma_{\rm ad}}$, 
revealing \textit{both} the equilibrium and SQ LL exponents.
A finite quasiparticle residue is retained during the quench, reflecting the Fermi gas nature of the initial state, getting suppressed gradually after the quench. 
The variety of quench induced phases offers a unique opportunity to design low dimensional correlated states on demand.



\begin{acknowledgments}

We thank A. Polkovnikov for stimulating comments.
B. D. thanks for the hospitality of MPIPKS in Dresden.  
This research has been  supported by the Hungarian Scientific 
Research Funds Nos.  K72613, K73361, CNK80991,
T\'{A}MOP-4.2.1/B-09/1/KMR-2010-0002  
and by the Bolyai program of the  Hungarian Academy of Sciences.
\end{acknowledgments}


\begin{widetext}

\section{Supplementary online material for "Crossover from adiabatic to sudden interaction quench in a Luttinger liquid"}

\section{Exact evaluation of the bosonic correlator of Eq. (9) in the main text for a linear quench}

For a linear quench, we can integrate Eq. (5) in the main text to obtain
\begin{gather}
v(q,t)=\frac{g_2(q)|q|}{2\omega^2(q)\tau}\left[\sin(\omega(q)t)-\omega(q)t\exp(-i\omega(q)t)\right]
\end{gather}
for $0<t<\tau$, and
\begin{gather}
v(q,t)=\frac{ig_2(q)|q|}{4\omega^2(q)\tau}\left[\exp(i\omega(q)(t-2\tau))-\exp(i\omega(q)t)+2i\omega(q)\tau\exp(-i\omega(q)t)\right]
\end{gather}
for $t>\tau$.

The exponent in the one-particle density matrix in
Eq. (9) in the main text is then evaluated in closed form for $t>\tau$,
$T=0$ and $L\rightarrow\infty$ as 

\begin{gather}
-\sum_{q>0}\left(\frac{2\pi}{qL}\right)4\sin^2\left(\frac{qx}{2}\right)|v(q,t)^2|=-\frac{g_2^2}{v^2}(I_1(\tau,x,R_0)+I_2(t,\tau,x,R_0)),
\end{gather}
where
\begin{gather}
I_1(\tau,x,R_0)=\int\limits_0^\infty dq \frac{\exp(-R_0q)}{q^3v^2\tau^2}\sin^2\left(\frac{qx}{2}\right)\left[\sin^2(\omega(q)\tau)+(\omega(q)\tau)^2\right],\\
I_2(t,\tau,x,R_0)=\int\limits_0^\infty dq \frac{\exp(-R_0q)}{q^2v\tau}\sin^2\left(\frac{qx}{2}\right)\left[-\sin(2\omega(q)t)+\sin(2\omega(q)(t-\tau))\right],
\end{gather}
which are evaluated as
\begin{gather}
I_1(\tau,x,R_0)=\frac{1}{32(v\tau)^2}\left[\sum_{r,s=\pm 1}\left\{\ln(2irv\tau +R_0+isx)(sx-iR_0+r2v\tau)^2\right\}+\right.\nonumber\\
+\left.2\sum_{s=\pm 1}\left\{\ln(R_0+isx)[(R_0+isx)^2+(2v\tau)^2]+\ln(2isv\tau +R_0)(R_0+2isv\tau)^2\right\}-4\ln(R_0)[R_0^2+(2v\tau)^2]\right]
\end{gather}
and
\begin{gather}
I_2(t,\tau,x,R_0)=\frac 12 \ln\left(R_0^2+4v^2(\tau-t )^2\right)-\sum_{s=\pm 1}\frac 14 \ln\left(R_0^2+(2v\tau-2vt -sx)^2\right)+\nonumber\\
+\frac{i}{8v\tau}\sum_{s=\pm 1}\left[2(2ivt-sR_0)\ln\left(\frac{sR_0+2iv(\tau-t)}{sR_0-2ivt}\right)+
\sum_{r=\pm 1}(rR_0-2ivt+isx)\ln\left(\frac{rR_0+i(2v\tau-2vt +sx)}{rR_0-i(2vt-sx)}\right)\right].
\end{gather}
The $0<t<\tau$ correlator is obtained as
\begin{gather}
-\sum_{q>0}\left(\frac{2\pi}{qL}\right)4\sin^2\left(\frac{qx}{2}\right)|v(q,t)^2|=-\left(\frac{g_2t}{v\tau}\right)^2\left(I_1(t,x,R_0)+I_2(t,t,x,R_0)\right).
\end{gather}
Expanding these in various limits are used to obtain the results cited in the paper, and their general $(t,\tau,x,R_0)$ dependence is used 
to generate Fig. 1. in the main text.

\section{Evaluation of the asymptotics of bosonic correlator of Eq. (9) in the main text for general  quench protocol}

In this section, we demonstrate that the asymptotic behavior what we obtained for a linear quench is universal, and arbitrary quench protocols lead to the same behaviour.
For arbitrary quench protocol $Q(t)$, the exponent in the one-particle fermionic density matrix of Eq. (9) in the main text at $T=0$ and $L\rightarrow\infty$ can be rewritten as
\begin{gather}
I=-\sum_{q>0}\left(\frac{2\pi}{qL}\right)4\sin^2\left(\frac{qx}{2}\right)|v(q,t)^2|=
-\frac{g_2^2}{4v^2}\int\limits_0^tdt_1\int\limits_0^tdt_2Q(t_1)Q(t_2){\partial_{t_1}}{\partial_{t_2}}f(t_1-t_2),
\end{gather}
where 
\begin{gather}
f(t)=\ln\left(1+\frac{x^2}{(R_0-2ivt)^2}\right)
\end{gather}
After partial integrations, it takes the form
\begin{gather}
I=-\frac{g_2^2}{4v^2}\left(Q^2(t)f(0)-2Q(t)\int\limits_0^{t}dt_1Q'(t_1)\textmd{Re}f(t_1-t)
+\int\limits_0^{t}dt_1Q'(t_1)\int\limits_0^{t}dt_2Q'(t_2)f(t_1-t_2)
\right).
\label{exp1}
\end{gather}
Here, by using $Q(t>\tau)=1$, the upper limit of integration would reduce to $\min\{t,\tau\}$. However, our considerations remain valid for smooth 
quench functions as well, reaching 1 only asymptotically, i.e 
$Q(t\gg\tau)\rightarrow 1$. 
Let us first consider the properties of the steady state in the limit of $t\gg (\tau,x/v)$, when the middle term in Eq. \eqref{exp1} does not contribute, 
and $Q(t\gg\tau)\cong 1$. 
For $t\gg |x/v|\gg\tau$, the first term yields $2\ln(|x|/R_0)$, while the last integral produces similar spatial decay and the adiabatic enhancement as
$2\ln(|x|/2v\tau)$. The exponent is
\begin{gather}
I=-\frac{g_2^2}{v^2}\ln\left|\frac{x}{\sqrt{R_02v\tau}}\right| \textmd{ for } t\gg |x/v|\gg\tau.
\end{gather}
In the $t\gg \tau\gg |x/v|$ limit, the last term also vanishes, and we are left with the adiabatic exponent
\begin{gather}
I=-\frac{g_2^2}{2v^2}\ln\left|\frac{x}{R_0}\right|  \textmd{ for } t\gg \tau\gg |x/v|.
\end{gather}

The other limit of interest is $|x| \gg (v\tau,vt)$, when the exponent simplifies in Eq. \eqref{exp1} with $f(t)$ replaced by  $f_1(t)=-2\ln(R_0-2ivt)$.
The first term always only contributes with a constant, $-2\ln(R_0)$.
In the limit of $|x| \gg vt\gg v\tau$, $Q(t)=1$, the third term becomes independent of both $x$ and $t$ and gives rise to the 
adiabatic enhancement factor as $-2\ln(\tau)$, 
and only the second term determines the temporal decay. The exponent is obtained as
\begin{gather}
I=-\frac{g_2^2}{v^2}\ln\left(\frac{t}{\sqrt{\tau_0\tau}}\right)   \textmd{ for } |x| \gg vt\gg v\tau
\end{gather}
In the limit of $|x| \gg v\tau\gg vt$, i.e. during the quench, the time dependence of $Q(t)$ is essential. 
The second term yields $4Q^2(t)\ln(t)$ to leading order, while the third term gives $-1/2$ times the second
term. Altogether, the exponent reads as
\begin{gather}
I=-\frac{g_2^2}{2v^2}Q^2(t)\ln\left(\frac{t}{\tau_0}\right)  \textmd{ for } |x| \gg v\tau\gg vt.
\end{gather}
These agree with the asymptotic expansion of the exact results for a linear quench.

\end{widetext}

\end{document}